\documentstyle[prl,aps,epsf]{revtex}
\begin{document}
\draft

\twocolumn[
\hsize\textwidth\columnwidth\hsize\csname@twocolumnfalse\endcsname\title
{Atomic-Scale Dynamics of the Formation and Dissolution \\
 of Carbon Clusters in SiO$_2$}
      
\author {Sanwu~Wang,$^{1,*}$ Massimiliano~Di~Ventra,$^{1,\dagger}$
S.~G.~Kim,$^{1,\ddagger}$ and Sokrates~T.~Pantelides$^{1,2,\S}$}

\address{$^1$Department of Physics and Astronomy,
         Vanderbilt University, Nashville, Tennessee 37235\\
         $^2$Solid State Division,
         Oak Ridge National Laboratory, Oak Ridge, Tennessee 37831}

\maketitle

\begin{abstract}

Oxidation of SiC produces SiO$_2$ while CO is released. A `reoxidation'
step at lower temperatures is, however, necessary to produce high-quality
SiO$_2$. This step is believed to cleanse the oxide of residual C without
further oxidation of the SiC substrate. We report first-principles
calculations that describe the nucleation and growth of O-deficient C
clusters in SiO$_2$ under oxidation conditions, fed by the production of
CO at the advancing interface, and their gradual dissolution by the supply
of O under reoxidation conditions. We predict that both CO and CO$_2$ are
released during both steps.

\end{abstract}

]

The most significant property of semiconductors is their ability to
sustain heterogeneous $n$-type and $p$-type doping. This property,
however, is eroded by high temperatures and high voltages that cause
intrinsic excitation of electron-hole pairs across the band gap. As a
result, semiconductors with significantly larger band gaps than silicon
have been investigated as candidates for electronic devices suitable for
high temperatures and high voltages. Silicon carbide is a particularly
attractive candidate because its native oxide is SiO$_2$ which works so
well as a dielectric in Si-based microelectronics. The presence of a
third element, however, namely C, results in a wide range of phenomena
that do not occur in the Si-SiO$_2$ system. In particular, oxidation of SiC
entails the production of CO which effuses through the oxide
\cite{Tan,Ventra}. Afanas'ev {\it et al.} have suggested that carbon
clusters at and near the interface form during oxidation
\cite{Afanasev,Bassler}, but the structure and dynamics of these clusters 
has not been established. Lipkin and Palmour found that, after oxidation, a
`reoxidation' step is necessary to produce high-quality oxides and
SiC-SiO$_2$ interfaces \cite{Lipkin,Chung}. During this step, oxygen is
supplied as during oxidation, but the temperature is lowered so that no
further oxidation takes place. In contrast, post-oxidation heat treatment
without the supply of O leads to an increase of charged defects in the
oxide \cite{Shenoy}. It is believed that the `reoxidation' step cleanses
the interface and bulk SiO$_2$ of residual carbon \cite{Afanasev}. Duscher
{\it et al.} recently presented direct experimental evidence for the
existence of carbon in as-grown samples and its removal after reoxidation
\cite{Duscher}.

The nucleation and growth of impurity clusters in semiconductors is a
generic problem for which total-energy calculations are well suited to 
provide detailed information. In this Letter we present the results of 
extensive first-principles
density-functional calculations that allow us to give a detailed account
of the nucleation and growth of O-deficient carbon clusters in SiO$_2$
during oxidation conditions and their dissolution during reoxidation
conditions. Basically, a CO molecule, generated at the advancing interface
and diffusing through the oxide, can bind weakly to an O site in the
SiO$_2$ network. A second CO molecule, however, can bind to the first and
the new complex is very stable. Additional CO molecules can extend the
cluster. {\it The process is helped by a passing CO that takes an O atom
from the cluster and effuses as CO$_2$.} The net result is O-deficient
clusters. During `reoxidation' at reduced temperatures, no further
oxidation of the SiC substrate occurs so that no CO is produced. Instead,
the supply of O atoms is used to `oxidize the C clusters'. The process
leads to the gradual dissolution of the C clusters through the formation
of both CO and CO$_2$. These results account for the available data and
lead to a prediction that can be tested, namely the effusion of CO$_2$
during both oxidation and reoxidation.
Note in particular that, during oxidation, there is simultaneous 
reoxidation, accounting for the fact that the concentration of C clusters 
is not excessive and that substantial amounts of CO are in fact released.
In contrast, during reoxidation, only dissolution of the clusters occurs. 

The objective of the calculations was to determine one or more sequences 
of steps that lead to the formation of O-deficient C clusters by the
aggregation of CO molecules and simultaneous release of O. Thus, sequences
of stable configurations were determined. Determination of the barriers 
for the formation of theses configurations is a laborious task because of
the multitude of atoms that are involved. Test calculations were carried
out in several cases by assuming specific pathways. Typical barriers were
smaller than 1 eV. Since the calculations represent upper bounds and the
relevant temperatures are above $1000^\circ$C, the processes identified in
this paper do in fact occur.
    
The calculations were performed within density functional theory, using
the pseudopotential method and a plane wave basis set. The
exchange-correlation effects were treated with local density approximation
in the form of Ceperley and Alder \cite{Ceperley}. We adopted the
Vanderbilt ultrasoft pseudopotentials \cite{Vanderbilt}. A plane wave
energy cutoff of 40 Ry was used. All the structures were fully relaxed.
Similar to earlier work \cite{Boero}, the calculations for the structure
relaxation were conducted with one {\bf k} point at (0.5, 0.5, 0.5), while
the densities of states (DOS) were calculated with 14 inequivalent {\bf k}
points.

A supercell ($9.60811\times10.93795\times10.62372$ \AA$^3$) containing 72
atoms, generated by Tu {\it et al.} \cite{Tu} with a Monte Carlo
bond-switching method \cite{Wooten}, was used to model the amorphous
SiO$_2$. This model corresponds to a density of 2.15 g/cm$^3$ for the
amorphous SiO$_2$, close to the experimental value of 2.20 g/cm$^3$ at the
standard pressure. The calculated DOS was found to be in good agreement
with the experimental photoemission spectra \cite{Fischer} and previous
first-principles calculations \cite{Sarnthein,Huang}, which used different
structure models to simulate the amorphous SiO$_2$. The SiO$_2$ band gap
was calculated to be 5.6 eV, the same as in earlier calculations
\cite{Huang}.

\paragraph*{CO molecules in a perfect SiO$_2$ network.}

We assume that the advancing interface during oxidation emits CO
molecules, as found in Refs. \cite{Tan} and \cite{Ventra}, and explore the
possible reactions that CO molecules can undergo in the SiO$_2$ matrix. A
perfectly bonded network without any defects is first considered. We find
that a single CO molecule can bind to a network O atom. The bound
configuration (hereafter [CO]) is shown in Fig. 1(a). There is a Si-C-O-Si
bridge with the extra O atom attached to the C atom so that the integrity
of the CO molecule is preserved. The bondlength of the CO molecule is now
1.21 \AA, 0.07 {\AA} longer than the bondlength of a free CO molecule at
an open interstitial site in SiO$_2$ (hereafter CO$^{(free)}$). While this
C-O bond is more likely a double-bond, another C-O bond between the carbon
and the original bulk oxygen atom appears to be a single-bond (with a
bondlength of 1.36 \AA). The total energy of the [CO] configuration is
0.23 eV lower than that of the CO$^{(free)}$. This is a rather small
binding energy, suggesting that both structures can exist. The energy cost
for the release of a CO molecule from SiO$_2$ is calculated to be 0.18 eV.
Hence, CO molecules at the open interstitial sites can easily escape from 
SiO$_2$, as found by experiments \cite{Tan}.

The binding energy of a CO molecule in the [CO] configuration, though
small, is sufficient to result in a substantial fraction of CO molecules
being bound at any given time [about 9 out of 10, as estimated from
$exp(-\Delta E/kT)$ with $\Delta E = 0.23$ eV and $kT = 0.1$ eV]. When a
diffusing CO molecule encounters a [CO], it can be attached to form a
two-molecule complex, [C$_2$O$_2$], {\it via} the reaction
\begin{equation}
[CO]+CO^{(free)} \to [C_2O_2].
\end{equation}
The energy gain is 1.6 eV, which suggests strong binding. The [C$_2$O$_2$]
configuration is shown in Fig. 1(b). The complex forms a bridge across a
network ring with the two halves of the bridge being on different planes.

When a third CO molecule gets near the [C$_2$O$_2$] structure, instead of
binding, it takes one of the oxygen atoms from the cluster and forms
either a CO$_2$ molecule at an open interstitial site (CO$_2^{(free)}$
configuration) or a CO$_2$-like bonded configuration ([CO$_2$], shown in
Fig. 1(c)). The [CO$_2$] configuration has essentially the same total
energy as CO$_2^{(free)}$ (only 0.04 eV lower, which is within the
uncertainty of the calculations). The remaining bonded configuration
([C$_2$O]) contains two carbon and one additional oxygen atoms, as shown
in Fig. 1(d). This structure is similar to the [C$_2$O$_2$] configuration
except for having one less oxygen atom. The above process responsible for
the formation of CO$_2$ is exothermic with a energy gain of 0.8 eV. Such a
reaction can be expressed as:
\begin{equation}
[C_2O_2]+CO^{(free)} \to [C_2O] + CO_2^{(free)} + 0.8 \ {\rm eV}.
\end{equation}

We conclude that oxidation of SiC results in the release of both CO and
CO$_2$. There are no experimental reports about CO$_2$ being released.
Special effort would be needed to detect it because it is probably present
in the chamber already.

As more CO molecules arrive, the [C$_2$O] complex undergoes additional
reactions. First, a CO molecule binds to the [C$_2$O] structure which
results in the formation of a [C$_3$O$_2$] configuration, shown in Fig.
2(a). The complex contains one C-O double bond and two C-O single bonds
with bondlengths of 1.15 {\AA} and 1.31 \AA, respectively. The bondlengths
of the two C-C bonds are 1.31 {\AA} and 1.37 \AA. The energy gain is again
very large, 2.0 eV.

The [C$_3$O$_2$] configuration is a very stable defect. The arrival of yet
another CO molecule does not lead to further growth. Though another CO can
be attached to [C$_3$O$_2$], there is an energy cost of 0.54 eV, which
makes the reaction highly unlikely. Similarly, a passing CO molecule
cannot take away an O atom from [C$_3$O$_2$] and become CO$_2$. The
process again involves energy cost (0.22 eV). If, however, by coincidence,
two CO molecules happen to arrive at a [C$_3$O$_2$] complex at the same
time (a rare event), the reaction
\begin{equation}
[C_3O_2]+2CO^{(free)} \to [C_4O_2]+CO_2^{(free)}
\end{equation}
can occur with an energy gain of 1.3 eV. The resulting [C$_4$O$_2$]
complex is shown in Fig. 2(b). Once it forms, this complex can again lose
an O atom to a passing CO, becoming a [C$_4$O] complex, shown in Fig.
2(c). The gain in energy is 1.4 eV.

The [C$_4$O] is unstable, however. Upon the arrival of a CO molecule, the
following reaction occurs
\begin{equation}
[C_4O]+CO^{(free)} \to [C_3O_{-2}] + 2CO_2^{(free)} + 1.1 \ {\rm eV}
\end{equation}
with an energy gain of 1.1 eV. Two CO$_2$ molecules are released and the
residual defect [C$_3$O$_{-2}$] consists of three C atoms in the space of
two O vacancies [Fig. 2(d)]. This last defect is stable against the
arrival of one or two CO molecules.

In summary, we have shown that {\it through a process of complex reactions
CO molecules passing through SiO$_2$ will form oxygen-deficient carbon
complexes through the emission of CO$_2$ molecules}. The most stable
defects are [C$_3$O$_2$] and [C$_3$O$_{-2}$]. The formation of the latter,
however, necessitates an intermediate step that requires the simultaneous
arrival of two CO molecules.

\paragraph*{CO molecules in SiO$_2$ with O vacancies.}

When oxygen vacancies exist in bulk SiO$_2$, a CO molecule fills a vacancy
by forming two Si-C bonds. The resulting configuration is shown in Fig.
1(e) (thereafter as [C]). The C-O bond is a double-bond with a bondlength
of 1.23 \AA. The binding energy of the CO molecule in the [C] structure
(relative to the molecule in the CO$^{(free)}$ structure) is 2.2 eV.
However, this structure is not stable. When an additional CO molecule
arrives,  the [C] structure transforms into the [C$_2$O] structure [Fig.
1(d)]. The energy gain is 3.0 eV.

\paragraph*{Reoxidation.}

We now examine the reactions that can occur during reoxidation when no new
CO molecules are produced but oxygen is available.

First, defects containing a single carbon atom such as [C] and [CO] can be
removed in the form of CO$_2$ molecules through the following processes:
\begin{eqnarray}
{[C] + O} & \to & [CO] + \ 8.8 \ {\rm eV,}\\
{[CO] + O} & \to & CO_2^{(free)} + 8.1 \ {\rm eV.}
\end{eqnarray}

Second, clusters of two carbon atoms, ie., those in the [C$_2$O] and
[C$_2$O$_2$] configurations can be dissociated and released in the form of
CO and CO$_2$ molecules as follows,
\begin{eqnarray}
{[C_2O] + O} & \to & [C_2O_2] + 7.6 \ {\rm eV,}\\
{[C_2O_2] + O} & \to & CO^{(free)} + CO_2^{(free)} + 6.6 \ {\rm eV.}
\end{eqnarray}

Similarly, the three-carbon clusters will be removed through the following
reactions:
\begin{eqnarray}
{[C_3O_{-2}] + O} & \to & [C_3O_{-1}] + 6.7 \ {\rm eV,}\\
{[C_3O_{-1}] + O} & \to & [C_3] + 6.8 \ {\rm eV,}\\
{[C_3] + O} & \to & [C_3O] + 7.5 \ {\rm eV,}\\
{[C_3O] + O} & \to & [C_3O_2] + 8.6 \ {\rm eV,}\\
{[C_3O_2] + O} & \to & [C_2O] + CO_2^{(free)} + 6.3 \ {\rm eV.}
\end{eqnarray}

Finally, the clusters with four carbon atoms will lose one carbon atom and
reduce to three-carbon clusters as follows,
\begin{eqnarray}
{[C_4O] + O} & \to & [C_4O_2] + 7.0 \ {\rm eV,}\\
{[C_4O_2] + O} & \to & [C_3O_2] + CO^{(free)} + 7.0 \ {\rm eV.}
\end{eqnarray}
The C$_3$ clusters will then be removed with further reoxidation, as shown
in Equs. (13), (7), and (8).

In summary, we have shown that the CO molecules that are generated during
oxidation of SiC diffuse through the network and can form a large variety
of complexes containing one, two, three, and four C atoms. One key result
is that these defects are oxygen deficient, as O atoms are removed by
combining with CO molecules forming CO$_2$ molecules. This phenomenon
provides a natural explanation of the `reoxidation' step since we also
showed that a supply of oxygen can dissolve the C clusters.

We would like to thank G. Duscher and L.~C. Feldman for useful
discussions. This work has been supported by the Defense Advanced Research
Program Agency grant MDA972-98-1-0007, the Electrical Power Research
Institute grant WO806905, NSF grant DMR-9803768, the Division of Materials
Sciences, US Department of Energy under contract DE-AC05-960R22464, and
the William A. and Nancy F. McMinn Endowment at Vanderbilt University.

\begin{figure}
\caption{Schematics of several species containing one and two C atoms:
(a) [CO]; (b) [C$_2$O$_2$]; (c) [CO$_2$]; (d) [C$_2$O]; and (e) [C]
configurations. The dark circles denote C atoms. The larger and smaller
grey circles are for Si and O atoms, respectively.}
\end{figure}

\begin{figure}
\caption{The minimum-energy structures corresponding to the configurations
containing clusters with three and four C atoms in SiO$_2$: (a)
[C$_3$O$_2$]; (b) [C$_4$O$_2$]; (c) [C$_4$O]; and (d) [C$_3$O$_{-2}$].}
\end{figure}

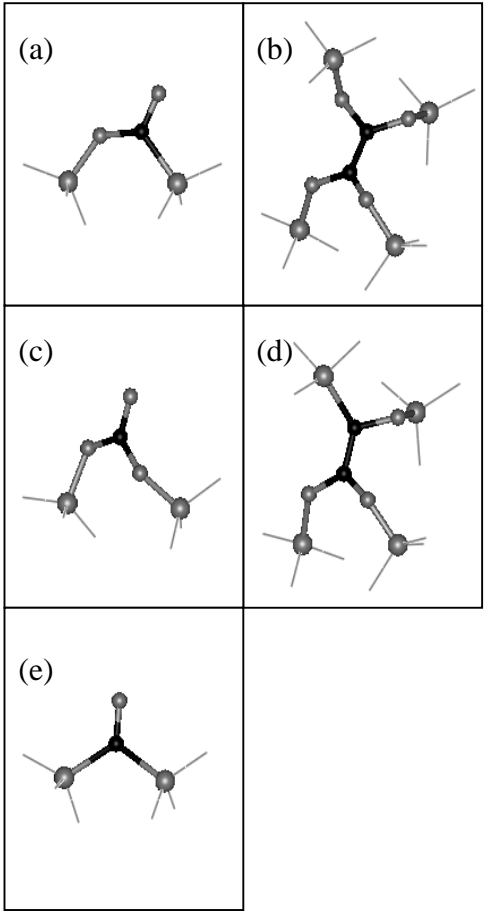
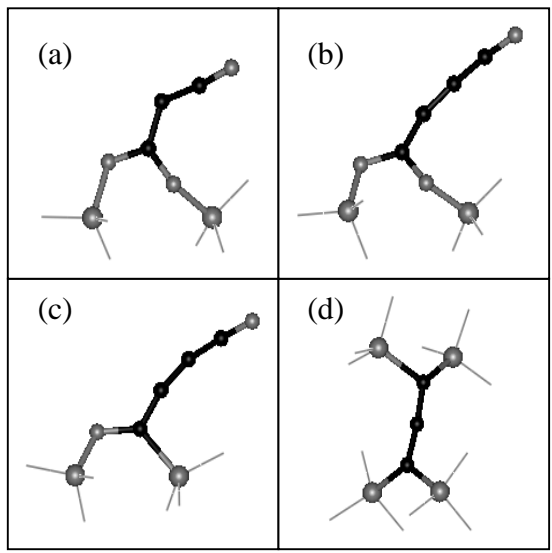
\end{document}